\documentclass{IOS-Book-Article}

\usepackage{times}
\normalfont
\usepackage[T1]{fontenc}
\usepackage{amsmath}
\usepackage{graphicx}

\begin{document}
\begin{frontmatter}                           

\title{Atomic quantum memories for light}

\author[A]{\fnms{Alexandra S.} \snm{Sheremet}},
\author[B]{\fnms{Oxana S.} \snm{Mishina}},
\author[B]{\fnms{Elisabeth} \snm{Giacobino}}\\
and
\author[A]{\fnms{Dmitriy V.} \snm{Kupriyanov}%
\thanks{Correspondence to: D.V. Kupriyanov Politehnicheskaya St. 29, 195251,
Saint-Petersburg, Russia. E-mail: kupr@dk11578.spb.edu.}}

\runningauthor{D.V. Kupriyanov et al.}
\address[A]{Department of Theoretical Physics, State
Polytechnic University\\ 195251, St.-Petersburg, Russia}
\address[B]{Laboratoire Kastler Brossel, Université Pierre et Marie Curie,
Ecole Normale Supérieure, CNRS, Case 74, 4 place Jussieu, 75252 Paris Cedex 05, France }

\begin{abstract}
We consider the coherent stimulated Raman process developing in an optically
dense and disordered atomic medium in application to the quantum memory scheme.
Our theoretical model predicts that the hyperfine interaction in the excited
state of alkali atoms can positively affect on the quantum memory efficiency.
Based on the concept of the coherent information transfer we analyze and
compare the memory requirements for storage of single photon and macroscopic
multi-photon light pulses.
\end{abstract}

\begin{keyword}
Quantum memory\sep Autler-Townes structure with hyperfine interaction\sep
ultracold atomic ensembles\sep coherent information
\end{keyword}
\end{frontmatter}

\thispagestyle{empty}
\pagestyle{empty}

\section*{Introduction}

Many implementations of quantum information processing, quantum computing or
secure communication need quantum memories for light. Various physical systems
are intensively studied as candidates for efficient light storage and
retrieval. Among those are the alkali atomic gases at room temperature
Refs.\cite{DengKoz}-\cite{HetetLam} and ultracold atomic ensembles
Refs.\cite{CDLK}-\cite{ChanKuz}, which have been successfully used for
demonstrating the quantum memory effect. Theoretical investigations of the
memory properties of atomic systems driven by a strong control field propose
reliable storage and retrieval of the quantum light in optically thick and
ultracold gases Refs.\cite{Gorshkov}-\cite{NunnJak}, and the predicted
parameters and conditions are experimentally feasible. In the present report we
discuss the role of the hyperfine interaction in alkali atoms and follow how it
modifies the stimulated Raman process applied to the quantum memory. We show
that the multilevel hyperfine structure of alkali atoms in their excited states
cannot be ignored and it is very important for the correct description of the
scattering process.

Having a quantum memory channel with certain efficiency it is important to
overcome a classical benchmark justifying that the memory is indeed quantum.
The fidelity of storage and retrieval is usually exploited to find such a
benchmark. The fidelity based benchmark has only been established for a limited
number of the quantum states and reveals to be different for each state, see
Refs.\cite{MasPopesku}-\cite{WolfPolzik},\cite{MKMP}. In the present report we
explore the benchmark based on the properties of the coherent information
originally proposed by Nielsen and Schumacher in Ref.\cite{SchumacherNielsen}.
This measure gives a convenient criterion if the memory channel preserves the
quantum correlations, which is intrinsically independent on the quantum state
nature. We apply it for evaluating the memory quality for different quantum
states of light.

\section{Quantum memory in a multilevel atomic system}

In this section we consider storage of a signal pulse in an ensemble of atoms
dressed by interaction with the off-resonant control field in the excitation
geometry shown in figure \ref{fig1}. In contrast with the widely used approach
of a single $\Lambda$-scheme we keep in our consideration both the interaction
with the control field and the hyperfine interaction in the exited state of the
alkali atom. As we show the interaction of the probe light with the medium,
which is determined by the dielectric susceptibility, is strongly modified by
the presence of the hyperfine interaction. We demonstrate that the storage and
retrieval of coherent signal light pulse with the rectangular temporal profile
can be optimized by relevant detuning of the control mode.

\subsection{Autler-Townes effect in the D${}_1$-line of alkali atom}

\begin{figure}[t]
\includegraphics{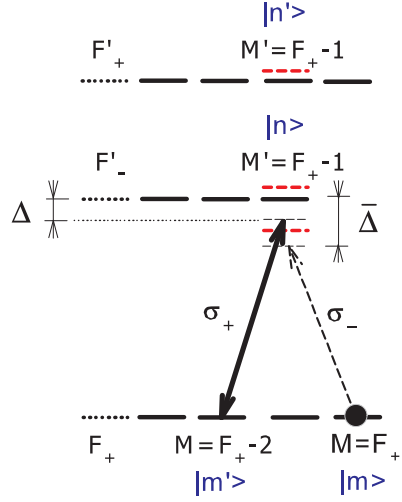}%
\caption{Schematic diagram showing the energy structure and the excitation channels
considered in D${}_1$ line of alkali atom. The atoms populate the upper hyperfine sublevel
of their ground state with maximal spin projection $F_{+}=I+1/2, M=F_{+}$. The system is
"dressed" by a strong mode of the control field in right-handed ($\sigma_+$) polarization with
frequency detuning $\Delta$ and probed by a weak mode in left-handed ($\sigma_{-}$) polarization.
The probe mode with frequency detuning $\bar{\Delta}$ scans the quasi-energy structure
of the Autler-Townes resonances. Locations of these resonances are shown by red
dashed bar lines.}
\label{fig1}%
\end{figure}

Let us consider D$_1$-line  of alkali atoms, which is typical for experiments
on quantum memories. As shown in figure \ref{fig1} all the atoms populate the
Zeeman sublevel with the maximal projection of the total angular momentum in
the ground state $F_{+}=I+1/2, M=F_{+}$ (where $I$ is atomic nuclear spin).
This state we denote as $|m\rangle$. Atomic response for the left-hand circular
polarized ($\sigma_-$) probe mode is controlled by the presence of the
right-hand circular polarized ($\sigma_+$) strong field. The energy levels of
the atom ($|n\rangle$: $F'_{-}=I-1/2, M'=F_{+}-1$ and $|n'\rangle$:
$F'_{+}=I+1/2, M'=F_{+}-1$) and the field mode, which is occupied by the strong
control field, transform to the quasi-energy resonance structure in accordance
with the Autler-Townes (AT) effect, see Refs.\cite{AutlerTownes,LethChebt}. For
the off-resonant control field two of the AT-resonances will be located near
the unperturbed atomic exited levels and one of the resonances will be located
near the frequency of the control field $\bar{\Delta}\sim \Delta$. The latter
resonance plays a crucial role for slowing down and storing the signal light
pulse. In comparison with widely used theoretical models including only one
exited state, see Refs.\cite{Gorshkov,HamSorPol}, our model shows that both the
exited states contribute to the AT resonance structure and thus should be
equally taken into consideration, see Refs.\cite{DengHagley,SherKupr}.
\begin{figure}[t]
\includegraphics[width=0.49\linewidth]{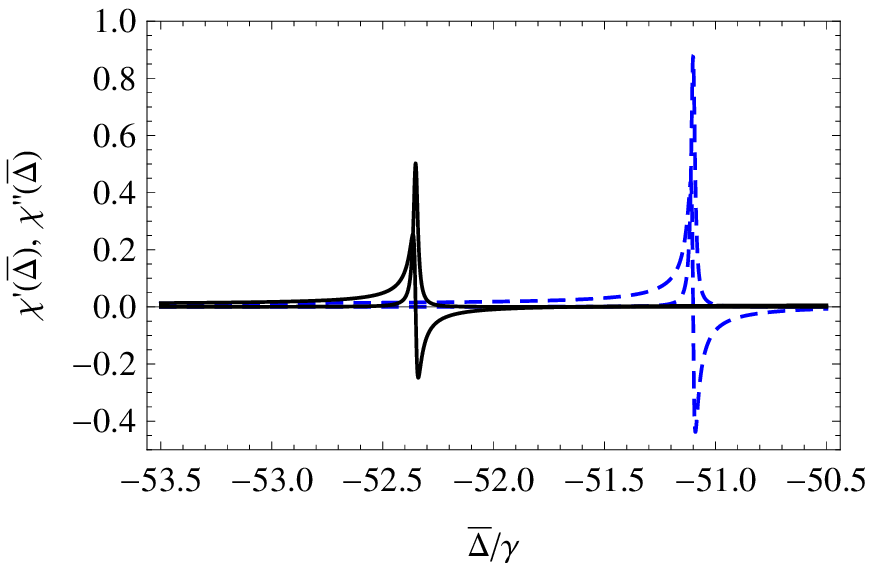}
\hfill
\includegraphics[width=0.49\linewidth]{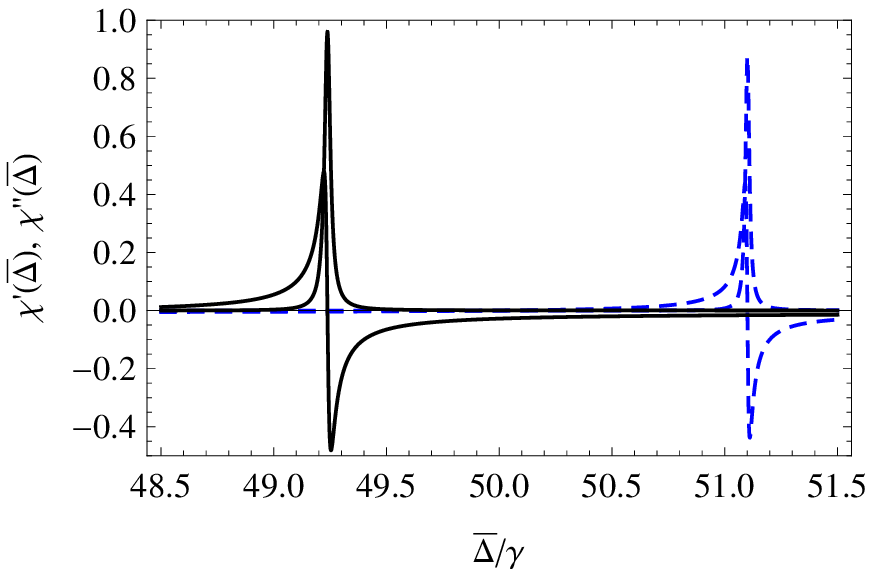}%
\caption{The absorption and dispersion parts of the sample dielectric susceptibility induced by
the control mode tuned by $\Delta=-50\gamma$ (left panel) and $\Delta=50\gamma$ (right panel) from the
atomic transition $|m'\rangle \to |n\rangle$, see figure \ref{fig1}, for the Rabi frequency
$\Omega_c=15\gamma$. The susceptibility components are scaled by $n_0(\lambda/2\pi)^3$
($n_0$ is the density of atoms and $\lambda$ is the transition wavelength). Black solid curves
relate to the exact calculations for the AT-triplet and the blue dotted curves indicate the dependencies
plotted in the $\Lambda$-scheme approximation.}
\label{fig2}%
\end{figure}

To demonstrate the importance of the hyperfine interaction in the upper state
and explain why it cannot be assumed as infinitely strong in figure \ref{fig2}
we present the spectral dependencies of the sample susceptibility
$\chi(\bar{\Delta})=\chi'(\bar{\Delta})+i\chi''(\bar{\Delta})$ for the probe
mode, which were calculated for the control field tuned to either red (left
panel) or blue (right panel) wings of the optical hyperfine transition
$|m'\rangle \to |n\rangle$, see figure \ref{fig1}. The Rabi-frequency
$\Omega_c$ is defined by the relevant matrix element for this transition. The
calculations were done for $D_1$-line of $^{133}$Cs atom, which has the
hyperfine splitting approximately equal to $256\gamma$, where $\gamma$ is the
atomic natural decay rate. The considered detunings $\Delta=\mp 50\gamma$ are
relatively small in comparison with the hyperfine splitting. However from the
plotted dependencies one can observe a significant deviation with predictions
of the three-level model, which presumes an infinitely strong hyperfine
interaction. The calculations based on three-level approximation with ignoring
the exited state $|n'\rangle$ are indicated by dashed curves. Our calculations
show that because of the hyperfine interaction the AT-resonance can be either
smaller (red wing) or bigger (blue wing) than predictions of the three level
model, when its amplitude does not depend on frequency detuning at all. It is
important that the enhancement of the AT-resonance takes place only for the
control mode tuned between the upper state hyperfine sublevels. This effect has
practical impact on the quantum memory since dispersion part of this resonance
is responsible for the delay and storage of the signal pulse in the atomic
system.

\subsection{Pulse storage and retrieval efficiency}

The advantage given by the enhancement of the dispersion properties of the
medium when the control field is tuned between the hyperfine states can be used
to provide longer delay of the signal pulse. The AT resonance shown in the
right panel of figure \ref{fig2} implements the delay of the signal pulse,
which spectral extension is essentially broader than the width of the
absorption part of this resonance. This allows to minimize the losses caused by
the incoherent scattering of the signal light to other directions out of the
coherent control mode. For optimal compromise between transparency and delay
the carrier frequency of the probe pulse and its duration should be properly
adjusted in respect to the AT resonance location and width and to the optical
depth of the sample.

\begin{figure}[t]
\includegraphics{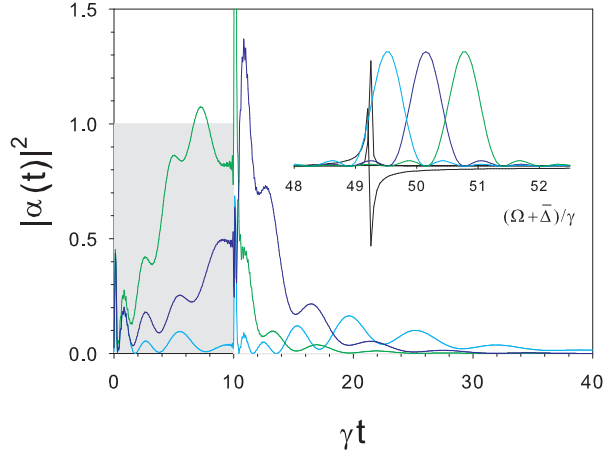}%
\caption{Time dependence of the probe pulses, with initially identical rectangular
profiles of duration $T=10\gamma^{-1}$ (shadowed box) and with different carrier frequencies, after they have
passed the optically thick sample with optical depth $n_0(\lambda/2\pi)^2L=50$. Inset shows the spectral
locations of the pulses with respect to the AT resonance. The central pulse has an optimal carrier
frequency to adjust the compromised balance between transparency and delay. The carrier
frequencies of two satellites are shifted by $\pm 2\pi/T$ from the central pulse frequency.}
\label{fig3}%
\end{figure}

In figure \ref{fig3} we show how three different signal pulses originally
having identical rectangular temporal profile but different carrier frequencies
would propagate the optically thick (for resonance radiation) sample. In inset
we show the spectral shape and position of the pulses as well as the location
of the AT resonance induced by the control field. The plotted dependencies show
how the original profiles of the signal pulses, given in dimensionless units by
$\alpha_{\mathrm{in}}(t)=\theta(t)-\theta(t-T)$, were $T$ is the pulse
duration, are modified after they have passed the optically thick medium. For
the round of calculations presented in figure \ref{fig3} we set optical depth
as $n_0(\lambda/2\pi)^2L\sim 50$. Our choice is motivated by the existing
experimental limits accomplished for ultracold atomic systems confined with
either magneto optic or quasistatic atomic traps.

The best compromise between transparency and delay of the probe is found for
the central carrier frequency among the presented in the inset of figure
\ref{fig3}. It is shifted from the induced resonance to minimize the absorption
but still overlap with the steep dispersion of the medium which provides the
pulse delay. For the pulse more overlapping the AT resonance (left spectrum)
the delay is longer but the pulse is mostly incoherently scattered. While
shifting the carrier frequency far away from the induced resonance (right
spectrum) allows to preserve the pulse shape as well as its integral intensity
but gives a quite short delay. That part of the signal pulse which emerges the
sample after the back front of the incoming pulse enters the sample at time $T$
could be stored by switching off the control field after time $T$. By switching
the control field on the stored signal pulse can be recovered with the same
shape as the part of dependence shown in figure \ref{fig3}, which follows after
time $T$. As was earlier verified in Ref.\cite{MLSSK} for off-resonant
stimulated Raman scattering the transient processes, associated with switching
on/off the control mode, only weakly interfere with the transport of the signal
pulse.

Detail optimization of the incoming pulse profile and use of the backward
retrieval scheme based on time reversion arguments could be further applied to
increase the memory efficiency. The strategic ideas of such optimization
procedure in example of three-level system have been discussed in Ref.
\cite{Gorshkov}. Here we just point out that the output profile of the
retrieved pulse, which is not rectangular any longer, determines the optimal
mode profile of the local oscillator field in the case of its homodyne
detection, i. e. if the quantum memory is considered in terms of continuous
variables. In the next section we briefly discuss the basic criteria for the
memory to be quantum and the required efficiency for the storage of different
quantum states.

\section{Figure of merit for the quantum memory channel}

Quality of the quantum memory is normally estimated via comparison with the
best possible classical strategy to memorize and reproduce the quantum state by
optimal measuring and reconstructing the state based on the measurement
results. The optimal classical protocol gives the lower bound for the
\textit{fidelity} of the quantum memory. Estimation of the classical fidelity
for the different quantum states requires different type of measurements, see
Refs.\cite{MasPopesku,MKMP,WolfPolzik}. As we shall point out here it is
possible to avoid the problem with classical fidelity benchmark and define
easier criterion of the memory quality for various of the quantum states. For
this goal we shall consider the stored light subsystem $Q$ as a part of
compound system, which is correlated i. e. entangled with the remaining
reference subsystem $R$. Such a situation seems typical for any quantum
network. Then in accordance with the general theory for noisy quantum
information channel developed by Schumacher and Nielsen in Ref.
\cite{SchumacherNielsen} we can define the \textit{coherent information}
transmitted through the memory channel. This parameter supposes to be positive
only if the quantum correlations between the retrieved light pulse and the
reference subsystem are preserved. In this case the memory could be qualified
as quantum since any possible measurement attacks would destroy the existing
quantum correlations.

In accordance with definition of Ref.\cite{SchumacherNielsen} the coherent
information transferred through the channel is given by the difference between
the von Neumann entropies of the retrieved subsystem $Q'$ and the entire system
$RQ'$ in the output of the channel
\begin{equation}
I_{e}=S_{Q'}-S_{RQ'}
\label{1}
\end{equation}
In input of the channel this quantity is equal to the entropy of the signal
light $S_Q$. If the channel losses were caused by interaction with environment,
originally existed in a vacuum state, then in accordance with the Schmidt
decomposition rule the subtracted term $S_{RQ'}$ would be equal to the entropy
transferred to environment $S_{E'}=S_{RQ'}$.

As an illustrating example let us consider first the memory channel when
$RQ$-system performs a pair of two polarization entangled photons, one of which
can be either stored and retrieved without changes or lost because of
incoherent scattering. With the verification scheme the presence of the
retrieved photon could be controlled in experiment and the experiment can be
repeated by a number of attempts till the single photon will appear in the
output. Such a conditional memory protocol is directly applicable to the
quantum repeater scheme aiming the qubit teleportation. The imperfection of the
conditional protocol is mainly determined by the thermal noise, which in our
situation can be induced by a random Raman photon spontaneously scattered into
the signal mode, and by low efficiency of the memory channel. Let us define by
$\eta$ the total memory efficiency including the spontaneous scattering and
verification losses and by $\mu$ the probability to extract in output the
random thermal photon. Then in the limit when $\mu\ll\eta<1$ the coherent
information passed through the channel is given by
\begin{equation}
I_e\sim 1 - \frac{3\mu(1-\eta)}{4\eta}\log_2\left[\frac{4e\eta}{\mu(1-\eta)}\right]%
\label{2}
\end{equation}
This quantity is positive and even approaches its maximal possible value under
the formulated conditions, which indicates that even at low level of the
quantum memory efficiency the quantum correlations can be preserved if the
excess noise is relatively small.

\begin{figure}[t]
\includegraphics[width=0.7\linewidth]{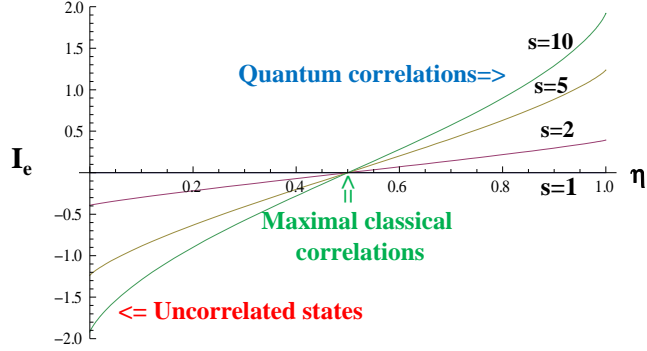}%
\caption{Coherent information representing the level of correlations between
two entangled light beams one of which has passed the memory channel with efficiency $\eta$.
The initial level of entanglement is given by the squeezing parameter $s$, see text.}
\label{fig4}%
\end{figure}

As a second illustrating example let us consider another typical situation when
the object $Q$ and reference $R$ subsystems are two initially identical
entangled macroscopic light beams. Such an EPR-type entangled state may be
produced either by combining two single-mode squeezed states at a beam splitter
or directly via a nonlinear two-mode squeezing interaction. For the sake of
convenience we prefer to discuss the former preparation scheme and parameterize
the state by its original level of squeezing $s\geq 1$ given by enhancement of
the square variance for the anti-squeezed quadrature at the input of beam
splitter. Then the average number of photons per mode for each beam at the
output of the beam splitter, which enters to the memory channel is given by
\begin{equation}
\bar{n}(s)=-\frac{1}{2}+\frac{1}{4}\left(s+\frac{1}{s}\right)%
\label{3}%
\end{equation}
The von Neumann entropy of each beam, which reproduces  the original volume of
the coherent information transmitted through the channel, is given
by
\begin{equation}
S\left(\bar{n}\right)=\left[\bar{n}+1\right]\ln\left[\bar{n}+1\right]%
-\bar{n}\ln\left[\bar{n}\right]%
\label{4}%
\end{equation}
Assuming that the environment was originally in a vacuum state and has absorbed
$(1-\eta)$ part of the beam, in the output of the channel the coherent
information reduces to
\begin{equation}
I_e=S\left(\eta \bar{n}\right)-S\left((1-\eta)\bar{n}\right)%
\label{5}%
\end{equation}
and it can be positive only if the level of losses expressed by efficiency
$\eta$ is smaller than 50\%. In figure \ref{fig4} we show the dependence of
$I_e=I_e(\eta)$ for different intensities of the input light beam i.e. for
different level of original squeezing $s$, see Eq.(\ref{3}). These dependencies
show that higher efficiency is required for storage of better correlated light
beams. Comparing the estimate (\ref{5}) with the conditional discrete variable
analog (\ref{2}) one can recognize that the unconditional continuous variable
scheme poses stronger requirements to the efficiency of the quantum memory and
thus to the quantum information processing in general.

\section{Conclusion}

In this paper the quantum memory protocol via stimulated Raman process in
D$_1$-line of alkali atom has been discussed in context of optimal choice of
the control mode and signal pulse parameters. The Raman process has been
considered beyond the widely used $\Lambda$-scheme approximation with paying
attention to the hyperfine interaction in the upper state. We have pointed out
that for the effective Raman-type storage of the signal pulse it is better to
tune the control field between the upper state hyperfine sublevels. The
parameters of the signal pulse, such as its carrier frequency and pulse
duration, should be properly adjusted in accordance with the Autler-Townes
resonance structure created by the control field.

In our estimate of the quality of the quantum memory channel we have followed
the criterium based on the volume of coherent information passed through the
channel. We have linked the coherent information with efficiency of the memory
channel and showed that the unconditional protocol requires higher efficiency
than it is in the case of conditionally verified single photon storage.

\section*{Acknowledgements}

This work was supported by RFBR (Grant No. 08-02-91355) and by the
Ile-de-France programme IFRAF. A.S. would like to acknowledge the financial
support from the charity Foundation "Dynasty."

\end{document}